\documentclass{article}
\usepackage[utf8]{inputenc}
\usepackage{mathtools}
\usepackage[table]{xcolor}
\usepackage{amssymb,graphicx}
\usepackage{epstopdf}
\usepackage{amsmath,amsfonts}
\usepackage{epsfig}
\usepackage{xcolor}
\usepackage[T1, T2A]{fontenc}
\usepackage[english]{babel}
\usepackage{float}
\usepackage{soul}
\usepackage{graphicx}
\usepackage{subcaption}
\usepackage{geometry}
\usepackage{hyperref}
\usepackage{psfrag}
\usepackage{slashed}
\usepackage{bm}
\usepackage{cite}
\usepackage[normalem]{ulem}

\usepackage{siunitx}
\usepackage{soul}
\begin{document}
\begin{flushright}
INR-TH-2024-016
\end{flushright}
\vspace{10pt}
\begin{center}
  {\LARGE \bf Non-renormalizable theories and finite formulation of QFT} \\
\vspace{20pt}
Y.~Ageeva$^{a}$\footnote[1]{{\bf email:}
    ageeva@inr.ac.ru}, P. Petrov$^{b}$\footnote[2]{{\bf email:}
    pavelkpetrov@mail.ru}, and
  M. Shaposhnikov$^{c}$\footnote[3]{{\bf email:} mikhail.shaposhnikov@epfl.ch}\\
\vspace{15pt}
  $^a$\textit{
Institute for Nuclear Research of
         the Russian Academy of Sciences,\\  60th October Anniversary
  Prospect, 7a, 117312 Moscow, Russia}\\
\vspace{5pt}
$^b$\textit{Cosmology, Gravity, and Astroparticle Physics Group,\\
Center for Theoretical Physics of the Universe,
Institute for Basic Science (IBS), Daejeon, 34126, Korea }\\
\vspace{5pt}
$^c$\textit{
 Institute of Physics, Ecole Polytechnique Federale de Lausanne, 
\\CH-1015 Lausanne, Switzerland
}
    \end{center}
    \vspace{5pt}
\begin{abstract}
In this paper, we show how the finite formulation of QFT based on  Callan-Symanzik equations can be generalised to the case of \textit{non-renormalizable}  theories. We derive an equation for  \textit{effective action} for an arbitrary single scalar field theory, allowing us to perform computations without running in intermediate divergencies. We illustrate the method with the use of $\lambda\phi^4 + \phi^6/M^2$ theory by the explicit (and fully finite) calculations of the effective potential as well as two-, four- and six-point correlation functions at one loop level and demonstrate that no quantum corrections to scalar mass $m^2$, depending on $M^2$-scale, are generated.

\end{abstract}
\newpage
\section{Introduction}
\label{sec:intro}

The papers~\cite{Mooij:2021ojy,Mooij:2021lbc,Mooij:2024rys} have shed light on a finite formulation of quantum field theory (QFT), which was proposed for the first time in Refs.~\cite{Blaer:1974foy,Callan:1975}  (as a proof of the validity of the multiplicative renormalisation scheme). This formulation delivers a divergence-free approach to renormalisation based on equations similar to the Callan-Symanzik (CS) equations. We call it the ``CS method'' throughout the text, following Refs.~\cite{Mooij:2021ojy,Mooij:2021lbc,Mooij:2024rys}. In these articles, it was shown that finite formulation of QFT perfectly works with the $\phi^4$ theory as well as with the case of several scalar fields: it is possible to calculate any correlation functions as well as any corrections to the effective potential in a fully finite way. The generalisation to the case of fermionic fields was worked out for QED in Ref.~\cite{Blaer:1974foy}.  It would seem that the next question is just around the corner: can the CS method work with the non-renormalizable theories? 

In this work, we show that such a generalisation, which can handle both renormalisable and non-renormalizable theories, indeed exists. To this end, we present the generalised CS equation, which is written through the effective action and can generate all possible Callan-Symanzik equations for $n$-point correlation functions as well as for effective potential in all orders of $\hbar$. 

Being equipped with such a generalisation, which can deal with the non-renormalizable theories, we explicitly calculate the one-loop correction to the effective potential and the \\$2$-, $4$-, and $6$-point correlation functions within some specific non-renormalizable theory. We do not face any divergences in the way: ingredients in the CS equations, intermediate calculations, and the results are finite. In considered non-renormalizable theory, we have two different energy scales: the $m$ mass of the scalar and some large (in comparison with $m$) scale $M$ associated with the operators of higher dimension. Our explicit calculations of both correlation functions and effective potential show that heavy-scale physics does not affect the $m^2$-order physics. Thus, we observe that no \textit{fine-tuning} (what is a sensitivity of physical observables to the variation of theory parameters) is required. This provides yet another argument in favour of the statement of Refs.~\cite{Mooij:2021ojy,Mooij:2021lbc,Mooij:2024rys} that the fine-tuning and naturalness problems (for original papers and different opinions see \cite{Gildener:1976ai,Weinberg:1975gm,Buras:1977yy,Susskind:1978ms,Mooij:2021ojy,Susskind:1982mw,Haber:1984rc,Dvali:1995,Martin:1997ns,Chung:2003fi,Giudice:2008bi,Feng:2013pwa,Dine:2015xga,Nath:2020xiz,Hebecker:2021egx,Park:2024kfn}) are related to the commonly used formalism of QFT based on divergent Feynman graphs and their multiplicative renormalisation, rather than representing a real physical challenge. 

The non-renormalisable field theories are most often considered as effective field theories, valid only below a certain energy. This is not necessarily the case, as these theories may be valid for arbitrary energies. The most notable example of this behaviour is associated with \textit{asymptotic safety} (for applications to gravity see \cite{Weinberg:GR,Reuter:1996cp}, and to the Standard Model ~\cite{Shaposhnikov:2009pv}. The analysis of non-renormalisable theories cannot be done perturbatively and requires some resummations (see, for instance, Ref.~\cite{Shaposhnikov:2023hrg}), most often done with the help of an exact functional renormalisation group (FRG) \cite{Wilson:1973jj,Polchinski:1983gv,Wetterich:1992yh}. In this paper, we have obtained a new exact equation for the effective action, which differs from FRG and so opens up a novel possibility to study non-renormalisable theories or EFTs.

This paper is organised as follows. We introduce the most general CS equation written in terms of some specific functional, which itself is connected to effective action in Sec.~\ref{sec:model}. In Sec.~\ref{sec:nonrenorm}, we illustrate how the CS method works with a simple non-renormalizable theory, including a higher-dimensional operator. To that end, in Sec.~\ref{sec:eff_poten}, we compute the one-loop correction to effective potential, while in Sec.~\ref{sec:green}, the corresponding correlation functions are found (in one loop as well).  We conclude in Sec.~\ref{sec:summary}. In Appendix \ref{app:monohar_compare},  we present the comparison of our results for beta functions with those of Ref.~\cite{Manohar:2024xbh}.

\section{Generalised Callan-Symanzik equation}
\label{sec:model}
In Refs.~\cite{Mooij:2021ojy,Mooij:2021lbc,Mooij:2024rys} it was shown that the Callan-Symanzik method works with the renormalisable theory of one or multiple massive scalar fields. A corresponding generalisation to fermion fields seems straightforward \cite{Blaer:1974foy}. Order by order (in $\hbar$ constant), one can recover the known results for $n$-point functions or the corrections to the effective potential, but in a manifestly finite way. However, the possibilities of the CS method do not end there. Let us show that the finite formulation of QFT can be extended to non-renormalisable theories and encoded in a unique equation that unifies the corresponding differential CS equations for both the renormalisable and non-renormalizable massive scalar theories.  

To clarify our further logic, let us begin with a brief review of the CS method for correlation functions \cite{Mooij:2021ojy,Mooij:2021lbc}. Take, for example, the $\phi^4$-theory with the Lagrangian
\begin{align}
\label{L_phi4}
    &\mathcal{L} = -\frac{1}{2}\partial_{\mu}\phi\partial^{\mu}\phi 
- \frac{m^2}{2}\phi^2 - \frac{\lambda}{4!}\phi^4.
\end{align}
There $m^2$ and $\lambda$ are introduced as finite parameters. The signature of the metric is $(-,+,+,+)$, which we use throughout the whole text. Within the CS method, the $\phi$'s $n$-point finite (renormalised, if the standard terminology is used) correlation functions $\Bar{\Gamma}^{(n)}$ (with overbar) are evaluated in a fully finite way by solving the following differential equations \cite{Mooij:2021ojy,Mooij:2021lbc}:
\begin{subequations}
\label{CS_for_phi4}
\begin{align}
    &2im^2G\Bar{\Gamma}^{(n)}_{\theta\theta} = \Big[\left(2m^2\frac{\partial}{\partial m^2}  + \beta \frac{\partial}{\partial   \lambda}\right)+n\gamma +\gamma_{\theta}\Big]\Bar{\Gamma}^{(n)}_{\theta},\\
    &2im^2G\Bar{\Gamma}^{(n)}_{\theta} = \Big[\left(2m^2\frac{\partial}{\partial m^2}  + \beta \frac{\partial}{\partial \lambda}\right)+n\gamma\Big]\Bar{\Gamma}^{(n)},
\end{align}
\end{subequations}
which are called \textit{Callan-Symanzik equations} (or just -- CS equations).\footnote{Note, that \eqref{CS_for_phi4} do not coincide with equations (3.4) from \cite{Mooij:2021ojy}. The reason is that we do not impose here the ``Callan boundary condition'' for the 2-point correlation function \cite{Callan:1975}
\begin{align*}
    \Bar{\Gamma}_{\theta}^{(2)}(k^2=0) = 1,
\end{align*} 
which leads to $G = (1+\gamma)$, see \cite{Mooij:2021ojy,Mooij:2021lbc}. Here $k$ 
is some external momentum. We will see that this condition is not necessary and that $G$ can be determined in a way that does not use it.} Different parameters which appear in \eqref{CS_for_phi4} are defined below.

The heart of the CS method is the meaning of the $\theta$ index there. This is so-called \textit{$\theta$-operation} and it is introduced as
\begin{align*}
    &\Gamma^{(n)}_{\theta} \equiv -i \times \frac{d}{dm_0^2}\Gamma^{(n)},\\
    &\Gamma^{(n)}_{\theta\theta} \equiv -i \times \frac{d}{dm_0^2}\Gamma^{(n)}_{\theta},
\end{align*} 
where $\Gamma^{(n)}$ are the bare Green's functions and $m_0$ is a bare mass \cite{Callan:1975,Mooij:2021ojy,Mooij:2021lbc}.\footnote{The bare quantities are only used at the derivation of the CS equations and never show up at any step of the computation of the finite Green's functions.}
The generalisation to the arbitrary $k$ theta-operations is as follows
\begin{equation*}
    \Gamma^{(n)}_{k\theta} \equiv 
-i \times \frac{d}{dm_0^2}\Gamma^{(n)}_{(k-1)\theta}\,,
\end{equation*}
where we introduce the shorthand notation $\Gamma^{(n)}_{k\theta}$, meaning $\Gamma^{(n)}_{1\theta} \equiv \Gamma^{(n)}_{\theta}$ for $k=1$, $\Gamma^{(n)}_{2\theta} \equiv \Gamma^{(n)}_{\theta\theta}$ for $k=2$, and etc.\footnote{The case of $k=1$ for the $\Gamma^{(n)}_{k-1,\theta}$ is nothing but $\Gamma^{(n)}_{0,\theta} \equiv \Gamma^{(n)}$, i.e. function without theta-operation.} The CS equation which connects the {\em finite} functions $\Bar{\Gamma}_{k\theta}^{(n)}$ to $\Bar{\Gamma}^{(n)}_{(k-1)\theta}$ is
\begin{align}
\label{k_theta}
2im^2G\Bar{\Gamma}_{k\theta}^{(n)} = \Big[\left(2m^2\frac{\partial}{\partial m^2} 
+   \beta \frac{\partial}{\partial \lambda}\right)+n\gamma+(k-1)\gamma_{\theta}\Big]\Bar{\Gamma}^{(n)}_{(k-1)\theta}.
\end{align}
The graphical representation of ${\Gamma}^{(n)}_{\theta}$ and ${\Gamma}^{(n)}_{\theta\theta}$ is related to the Feynman diagrams: $\theta$-operation splits every propagator into two parts by inserting a new kind of vertex, which we will denote as a cross in this paper, following \cite{Mooij:2021ojy,Mooij:2021lbc,Mooij:2024rys}. So, applying $\theta$-operation on a diagram with $n$ propagators returns $n$ new diagrams, each with $(n+1)$ propagators. One can ``heal''  the relevant UV-divergent bare diagrams (i.e. make them UV-convergent) with the use of a required (two $\theta$ operations are needed for $\Bar{\Gamma}^{(2)}$ and one for $\Bar{\Gamma}^{(4)}$ for the case of $\phi^4$ theory) number of theta-operations until the diagram becomes finite. The set of the requisite diagrams is determined with the use of the so-called ``skeleton'' expansion \cite{Callan:1975}. After that, finite expressions for $\Bar{\Gamma}^{(4)}_{\theta}$ and $\Bar{\Gamma}^{(2)}_{\theta\theta}$ should be fed to the CS equations. To compute the Greens functions with a larger number of legs, the skeleton expansion is to be used \cite{Callan:1975}. We also bear in mind that $G$ in \eqref{CS_for_phi4} is given by
\begin{align*}
    G\equiv \left[\frac{\partial m^2}{\partial m_0^2}\right]^{-1} Z_{\theta}, 
\end{align*}
and the object $Z_{\theta}$ was introduced to renormalize $\Gamma^{(n)}_{\theta}$ correlation function. The anomalous dimensions are given by
\begin{align*}
    \gamma \equiv m^2 \left[\frac{\partial m^2}{\partial m_0^2}\right]^{-1}\frac{\partial 
\;\text{ln}\;Z}{\partial m_0^2}, \quad \gamma_{\theta} \equiv 2m^2\left[\frac{\partial m^2}{\partial m_0^2}\right]^{-1}\frac{\partial \;\text{ln}\;Z_{\theta}}{\partial m_0^2},  
\end{align*}
where $Z$ renormalizes $\Gamma^{(n)}$ correlation function; for beta-function we have
\begin{align}
\label{our_beta_def}
    \beta \equiv 2m^2 \left[\frac{\partial m^2}{\partial m_0^2}\right]^{-1}\frac{\partial \lambda}{\partial m_0^2}.
\end{align}
All $G$, $\gamma$, $\gamma_{\theta}$, and $\beta$ can be found during the solution of \eqref{CS_for_phi4} (for example, with the use of boundary conditions), see Ref.~\cite{Mooij:2021ojy,Mooij:2021lbc} for the details. Defining $G$, $\gamma$, $\gamma_{\theta}$, and $\beta$ as well as finite $\Bar{\Gamma}^{(n)}_{\theta}$ and $\Bar{\Gamma}^{(n)}_{\theta\theta}$, the equations \eqref{CS_for_phi4} now can be solved to find $\Bar{\Gamma}^{(n)}$. For example, it is enough to consider these two equations \eqref{CS_for_phi4} to find two- and four-point correlation functions at one loop level in the framework of \eqref{L_phi4}. This ends our review of how the CS method works for $n$-point functions.

For our purposes, the next step is to introduce the CS equations for \textit{effective action}. The latter is the generating functional for the strongly connected Green's functions:
\begin{subequations}
\label{action_VS_green}
\begin{align}
    &\Gamma_{\text{eff}} = \sum_n \frac{1}{n!} \int d^4x_1\ldots d^4x_n \Bar{\Gamma}^{(n)}(x_1\ldots x_n) \phi_0(x_1) \ldots \phi_0(x_n),\\
    &\Gamma_{\text{eff},\theta} = \sum_n \frac{1}{n!} \int d^4x_1\ldots d^4x_n \Bar{\Gamma}_{\theta}^{(n)}(x_1\ldots x_n) \phi_0(x_1) \ldots \phi_0(x_n),
\end{align}
\end{subequations}
etc; here $\phi_0$ denotes the classical background field,  see Ref.~\cite{Coleman:1973jx}. Using notations \eqref{action_VS_green} together with \eqref{CS_for_phi4}, within the theory \eqref{L_phi4}, one can immediately write down the CS equations for the effective action:
\begin{align}
\label{theta1}
    &2im^2G\Gamma_{\text{eff},\theta\theta} = \Big[\left(2m^2\frac{\partial}{\partial m^2} 
+ \beta \frac{\partial}{\partial \lambda} \right)
+\gamma \phi_0 \frac{\delta}{\delta \phi_0} + \gamma_{\theta} \Big]\Gamma_{\text{eff},\theta},\\
\label{theta0}
    &2im^2G\Gamma_{\text{eff},\theta} = \Big[\left(2m^2\frac{\partial}{\partial m^2}  + \beta \frac{\partial}{\partial \lambda}\right)+\gamma \phi_0 \frac{\delta}{\delta \phi_0}\Big]\Gamma_{\text{eff}},
\end{align}
respectively; here $\delta/\delta \phi_0$ is the functional derivative with respect to $\phi_0$.

As was shown in Ref.~\cite{Mooij:2024rys}, the Callan-Symanzik equations can be written for \textit{effective potential}. To this end, we briefly recall that the effective action can be written as an expansion in powers of derivatives, i.e.
\begin{align}
\label{deriv_expans}
    \Gamma_{\text{eff}} = - i \int d^4x [\Gamma(\phi_0) 
+ K(\phi_0) + \ldots],
\end{align}
where $\Gamma(\phi_0)$ is an effective potential and $K(\phi_0)\propto (\partial_{\mu}\phi_0)^2$  is an effective kinetic term. The effective potential is given by the sum of all Feynman diagrams with only external scalar lines and with vanishing external momenta. Thus, the expression \eqref{deriv_expans} together with \eqref{theta1}  and \eqref{theta0} lead to the CS equations for the effective potential, where $\phi_0$ is now a constant independent of a space-time point:
\begin{subequations}
\label{CS_V}
\begin{align}
\label{CS_V_2}
&2im^2G\Gamma_{\theta\theta} = \Big[\left(2m^2\frac{\partial}{\partial m^2} 
+ \beta \frac{\partial}{\partial \lambda} \right)+\gamma \phi_0 \frac{\partial}{\partial \phi_0} 
+ \gamma_{\theta} \Big]\Gamma_{\theta},\\
&2im^2G\Gamma_{\theta} = \Big[\left(2m^2\frac{\partial}{\partial m^2}  
+ \beta \frac{\partial}{\partial \lambda}\right)
+\gamma \phi_0 \frac{\partial}{\partial \phi_0}\Big]\Gamma\,.
\label{CS_V_1}
\end{align}
\end{subequations}

Above, we introduce all the needed ingredients and are ready to generalise the CS method to a non-renormalizable case. To this end, one has to account for the following points:
\begin{enumerate}
    \item The key point is that non-renormalizable theories may include different operators, each of a different dimension. Such operators produce diagrams with an arbitrarily high degree of UV divergence. However, this is not the problem for the CS method since one can apply as many theta operations as needed to make the relevant Feynman graphs convergent.
    \item Operators are always included in the Lagrangian together with corresponding coupling constants. This means that the generalisation of the CS equation will contain new beta functions related to these new coupling constants.
\end{enumerate}
That is why, taking into account the first (1) point for the above discussion, we introduce a functional:
\begin{align}
\label{functional}
    \Gamma_{\text{eff}}(\theta,\phi_0) 
= \sum_{n=0}^{\infty} \Gamma_{\text{eff},n\theta}(\phi_0)\frac{\theta^n}{n!},
\end{align}
where we use the shorthand notations $\Gamma_{\text{eff},1\theta} \equiv \Gamma_{\text{eff},\theta}$, $\Gamma_{\text{eff},2\theta} \equiv \Gamma_{\text{eff},\theta\theta}$, etc. The introduction of the functional \eqref{functional} immediately allows us to write an equation
\begin{align}
\label{general}
    &\Big[2m^2 \Big(\frac{\partial}{\partial m^2} - i G \frac{\partial}{\partial \theta }\Big) 
+ \gamma_{\theta} \theta \frac{\partial}{\partial \theta } 
+ \gamma \phi_0 \frac{\delta}{\delta \phi_0} + \sum_i \beta_i \frac{\partial}{\partial \lambda_i}\Big] \Gamma_{\text{eff}}(\theta,\phi_0) = 0,
\end{align}
which unifies all possible CS equations for effective action and manifests itself as 
\textit{a general CS equation} we are looking for. This equation represents a new result of this paper.  The Eq.~\eqref{general} is different from the exact renormalization group equations \cite{Wilson:1973jj,Polchinski:1983gv,Wetterich:1992yh}, and does not contain any reference to the energy ``cutoff'' used in FRG.  Introducing the term   
\begin{align*}
    \sum_i \beta_i \frac{\partial}{\partial \lambda_i},
\end{align*}
allows us to take into account the second (2) point from the discussion above. For example, in the case of $\lambda\phi^4$ theory, it is just given by
\begin{align*}
    \sum_i \beta_i \frac{\partial}{\partial \lambda_i} \to \beta \frac{\partial}{\partial \lambda},
\end{align*}
while in the case of $\lambda\phi^4$ plus some higher dimension operator $g\phi^6$ we  have
\begin{align*}
    \sum_i \beta_i \frac{\partial}{\partial \lambda_i} \to \beta \frac{\partial}{\partial \lambda} + \Omega_{g} \frac{\partial}{\partial g},
\end{align*}
where $\Omega_g$ is a beta-function for $g$ coupling constant. The equation for effective potential is the same as \eqref{general} but with the constant field $\phi_0$.

So, it does no matter now which theory is under consideration: renormalisable or non-renormalizable 
one with the set of operators of any dimension. If one considers a non-renormalizable 
theory with higher order dimension operators with corresponding coupling constants, then the CS 
equation \eqref{general} together with \eqref{functional} gives as many differential equations with arbitrary demanded number of theta-operations (to make the relevant graphs finite) as well as allows to determine all related beta-functions. 

As a result, now we have the generalisation of the CS method: step by step, with the equation \eqref{general}, one can recover any order (for example, by $\hbar$) corrections to effective action or potential as well as to any $n$-point correlation functions in a manifestly finite way. We have shown that the CS method may work even when one includes some higher dimension operators into the Lagrangian (this is precisely the case of non-renormalizable theories). For the latter, there are no problems: the eq.~\eqref{general} takes this into account, just adding new beta functions (related to these new operators) in all CS equations. 

Surely, non-renormalizable theory remains non-renormalizable in the CS approach. In the standard renormalisation schemes, we need an infinite number of counterterms to cancel all the infinities in these theories. The manifestation of the non-renormalisability of the theory in the CS method is the infinite amount of $\theta$ operations that are needed to make computations in all orders of perturbation theory and thus an infinite number of the integrations constants which determine the theory. Still, the perturbative expansion can be organised in a regular way, which is used in effective field theory description of non-renormalisable theories. Namely, in addition to $\hbar$ expansion counting the number of loops, one may use a specific order of the mass scale $M$ associated with the operators with a mass dimension greater than $5$. Within a specific order in $1/M$, the number of the necessary $\theta$ operations is finite, as well as the number of the integration constants, making the theory predictable. An example of the next Section clarifies how this procedure can be implemented.

\section{Callan-Symanzik method and scalar non-renormalizable theory}
\label{sec:nonrenorm}
In the previous chapter, we found the generalization of the CS method, which can work with the non-renormalizable theories.  To illustrate how this works, we proceed with the explicit evaluation of one-loop correction to the effective potential  and the correlation  functions in a simple non-renormalizable theory with the Lagrangian including all dimension six operators
\begin{align}
\label{six_dim_all}
    \mathcal{L} = &-\frac{1}{2}\partial_{\mu}\phi\partial^{\mu}\phi 
- \frac{m^2}{2}\phi^2 - \frac{\lambda}{4!}\phi^4 - \frac{g}{6!M^2} \phi^6 
+ \frac{\xi}{M^2} \phi (\Box^2 \phi)   + \frac{f}{3!M^2}\phi^3 \Box \phi ,
\end{align}
where $m^2$, $\lambda$, $g/M^2$, $\xi/M^2$, and $f/M^2$ are \textit{finite}, so all physical quantities are expressed as functions of these parameters. Here $M$ is some large (in comparison with $m$) parameter of mass dimension. 

Now, to find the Green's functions, eq. (\ref{general}) should be written with an account of all coupling constants of the theory, namely $m^2,\lambda,g,\xi$ and $f$. However, in one-loop approximation and the first order in $1/M^2$ it turns out that \eqref{six_dim_all} can be simplified with the use of \textit{reparametrisation freedom}. Indeed, considering the following field redefinition:
\begin{align}
\label{phi_transform}
    \phi \to \phi + C_1\frac{\phi^3}{M^2} + \frac{C_2 \Box \phi}{M^2} 
+ C_3 \frac{m^2 \phi}{M^2}\,,
\end{align}
it is possible to get rid of some terms (of order $1/M^2$) in \eqref{six_dim_all}. For us, the most convenient choice is to keep only the potential-like term $\sim \phi^6$. So, the following choice of constants
\begin{align*}
    &C_1 = - \frac{f}{3!} - \frac{\xi \lambda}{3!},\\
    &C_2 = C_3 =- \xi,
\end{align*}
brings us from \eqref{six_dim_all} right to the desired Lagrangian 
\begin{align}
\label{L}
    &\mathcal{L}_{\text{New}} =-\frac{1}{2}\partial_{\mu}\phi\partial^{\mu}\phi 
- \frac{1}{2}\tilde{m}^2\phi^2  - \frac{\tilde{\lambda}}{4!}\phi^4  - \frac{\tilde{g}}{6!M^2} \phi^6,
\end{align}
with
\begin{align*}
    \tilde{\lambda} = \lambda + 4!C_1 \frac{ m^2}{M^2} +4C_3 \frac{\lambda m^2 }{M^2},\\
    \tilde{m}^2 = m^2  + 2C_3 \frac{m^4}{M^2} ,\\
    \tilde{g} = g +120 C_1 \lambda   .
\end{align*} 
We can use the Lagrangian \eqref{L} in all our further calculations (we will omit tildes on $\tilde{m}$, $\tilde{\lambda}$, and $\tilde{g}$ and index ``New'' from \eqref{L}  everywhere in the text below in order not to encumber the formulas). It is important to stress, though, that the non-linear character of the transformation (\ref{phi_transform}) shows up in the higher loops, which leads to the necessity to keep all the coupling constants \cite{Manohar:2024xbh}.

Another important remark to make is as follows: though the found field redefinition \eqref{phi_transform}  helps to reduce the number of terms in \eqref{six_dim_all}, it is impossible to find a redefinition of the field to get rid of the terms with the derivatives in higher orders in $1/M^2$.  For example, if the dimension eight operators \cite{Henning:2017fpj} are included, the convenient (but not the unique) minimal choice is \cite{Henning:2017fpj}
\begin{align}
\label{eight}
    \mathcal{L}_8 \to \frac{\lambda_{8,1}\phi^8}{M^4} + \frac{\lambda_{8,2}\left[(\partial_{\mu}\phi)^2\right]^2}{M^4},
\end{align}
where $\lambda_{8,1}$ and $\lambda_{8,2}$ are some coupling constants.\footnote{Another example of how one can write the dimension eight operators is given in Ref.~\cite{Atance:1996yd}.}

Choosing the non-renormalizable theory \eqref{L}, in Sec.~\ref{sec:eff_poten}, we begin with the finite approach to computing the one-loop correction to effective potential keeping only the $1/M^2$ terms; in the Sec.~\ref{sec:green} we turn to calculation of two-, four- and six-point functions in the same theory with the Lagrangian \eqref{L} (in one loop approximation and $1/M^2$ order as well). 

\subsection{One loop correction to the effective potential}
\label{sec:eff_poten}
\begin{figure}
    \centering
\includegraphics[scale=0.43]{
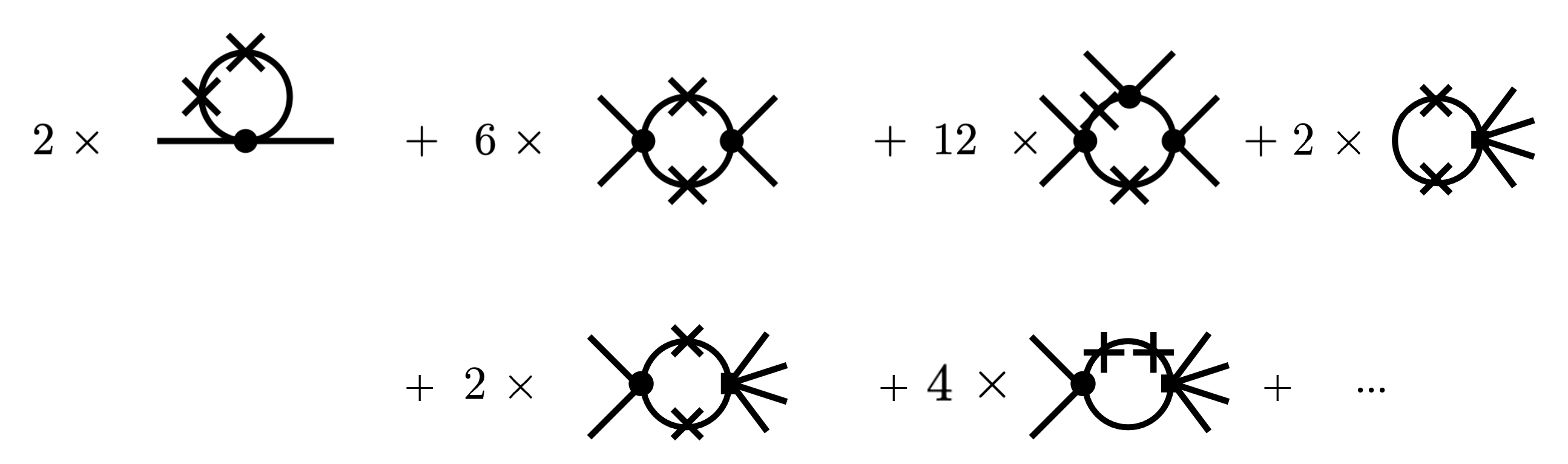}
    \caption{One loop contributions to $\Gamma_{\theta\theta,1}$. Square vertex corresponds 
to $ g\phi_0^4$ term.}
    \label{fig:phi_eff}
\end{figure}

Now, we get to the explicit calculation of one-loop correction to the effective potential. Firstly, we define (in the same manner as in Ref.~\cite{Mooij:2024rys}) the expansion
\begin{align*}
    \Gamma = \Gamma_0 + \hbar\cdot \phi_0^4 \cdot \Gamma_1(\phi_0^2/m^2) + \mathcal{O}(\hbar^2),
\end{align*}
where $\Gamma_0$ is the classical potential, which reads
\begin{align}
\label{Gamma0}
    \Gamma_0 = \frac{m^2\phi_0^2}{2} + \frac{\lambda\phi_0^4}{4!} + \frac{g\phi_0^6}{6! M^2},
\end{align}
and  $\Gamma_1$ is the one-loop correction, which we are going to find in this Section. We also define
\begin{align*}
    &\Gamma_{\theta} = \Gamma_{\theta,0} + \hbar \cdot \Gamma_{\theta,1} 
+ \mathcal{O}(\hbar^2),\\
    &\Gamma_{\theta\theta} = \Gamma_{\theta\theta,0} + \hbar \cdot \Gamma_{\theta\theta,1} 
+ \mathcal{O}(\hbar^2).
\end{align*}
In the theory \eqref{L} and in one loop approximation, it is necessary and sufficient to apply two $\theta$ operations on the diagrams to make them finite. All one-loop contributions are shown in Fig.~\ref{fig:phi_eff}. Indeed, for example, the first and the fourth diagrams in Fig.~\ref{fig:phi_eff} (upper line) have three propagators, so they are proportional to 
\begin{align*}
    \sim \int \frac{d^4l}{(2\pi)^4}\frac{1}{(l^2+m^2)^3},
\end{align*}
and these integrals are UV convergent. Other diagrams in Fig.~\ref{fig:phi_eff} converge even better since they include more propagators. In the calculations below, we also neglect contributions from vacuum energy. Next, we denote the corrections to all $\gamma$, $\gamma_{\theta}$, $\beta$ and $\Omega_g$ as:
\begin{align*}
    &G = G_0 + \hbar \cdot G_1,\\
    &\gamma = \gamma_0 + \hbar \cdot \gamma_1,\\
    &\gamma_{\theta} = \gamma_{\theta, 0} + \hbar \cdot \gamma_{\theta,1},\\
    &\beta =  \beta_0 + \hbar \cdot \beta_1,\\
    &\Omega_g = \Omega_{g,0} + \hbar \cdot \Omega_{g,1}.
\end{align*}
Let us show, how we find all zero order $G_0$, $\gamma_0$, $\gamma_{\theta,0}$, $\beta_0$ and $\Omega_{g,0}$ parameters. To that end, we consider the one-loop correction to the \textit{effective kinetic} term:
\begin{align}
\label{kinetic}
    &K = \frac{1}{2} (\partial_{\mu}\phi_0)^2 \Big\{K_0 
+ \hbar \cdot K_1\big(\phi_0^2/m^2,\lambda,g/M^2\big)+ \mathcal{O}(\hbar^2)\Big\},
\end{align}
and the corresponding CS equation, which can be obtained with the use of \eqref{deriv_expans} and \eqref{general}, reads:
\begin{align}
\label{CS_K}
 &2m^2iG K_{\theta} = \Big[\left(2m^2\frac{\partial}{\partial m^2}  
+ \sum_i \beta_i \frac{\partial}{\partial \lambda_i}\right)
+\gamma \phi_0 \frac{\partial}{\partial \phi_0}\Big]K,
\end{align}
where we also defined
\begin{align*}
    &K_{\theta} = \frac{1}{2} (\partial_{\mu}\phi_0)^2 \cdot\Big\{K_{\theta,0} 
+ \hbar \cdot K_{\theta,1} + \mathcal{O}(\hbar^2)\Big\}.
\end{align*}
At the tree level, we have
\begin{align*}
    K_0 = 1,\quad K_{\theta,0} = 0,
\end{align*}
so, evaluating CS equation \eqref{CS_K} at zeroth order in $\hbar$, we find out
\begin{align}
\label{g0}
    \gamma_0 = 0.
\end{align}
Other zeroth order parameters can be defined from the CS equations for the effective potential \eqref{CS_V}. At tree level order $\Gamma_0$ is given by \eqref{Gamma0}, and thus  
\eqref{CS_V_1} leads to
\begin{align*}
    \Gamma_{\theta,0} = - i \frac{\phi_0^2}{2},
\end{align*} 
where we also use the result \eqref{g0}. If we substitute $\Gamma_{\theta,0}$ back to the equation \eqref{CS_V_1}, then we arrive to (again in zeroth order in $\hbar$) 
\begin{align*}
    m^2 \phi_0^2 -G_0 m^2 \phi^2_0 + \frac{\beta_0 \phi^4_0 }{24} + \frac{\Omega_{g,0} \phi_0^6}{720M^2} = 0,
\end{align*}
which must be satisfied for any arbitrary $\phi_0$. Thus, it defines all the parameters as follows
\begin{align*}
    &G_0 = 1,\\
    &\beta_0 = 0,\\
    &\Omega_{g,0} = 0.
\end{align*}
Finally, we consider another CS equation \eqref{CS_V_2} in zeroth by $\hbar$ order; having $\Gamma_{\theta\theta,0} = 0$ it gives
\begin{align*}
    \gamma_{\theta, 0} = 0.
\end{align*}

Defining all the tree-level parameters, we turn to the CS  equations at $\hbar$ order; evaluating the equation for the effective kinetic term \eqref{CS_K} at this order, we arrive at:
\begin{align}
\label{CS_K_h}
    \gamma_1 - im^2 K_{\theta,1} - x \frac{\partial K_1}{\partial x} = 0,
\end{align}
where we omit some overall factors and introduce a 
new dimensionless variable 
\begin{align}
\label{x}
    x \equiv \frac{\phi_0^2}{m^2}.
\end{align}
The function $K_{\theta,1}$ is already finite and can be found, for example, from the direct use of the background field method together with an adiabatic expansion of effective action, i.e. counting the number of derivatives acting on field $\phi_0$. Let us bring the sketch of the latter method. Firstly, substitute $\phi \to \phi_0 + \delta \phi$ into \eqref{L} and write the quadratic by $\delta \phi$ part:
\begin{align*}
    \mathcal{L}^{(2)} = -\frac{1}{2} (\partial_{\mu}\delta \phi)^2 - \frac{m^2}{2}  (\delta \phi)^2 - \frac{\lambda}{4}\phi_0^2\delta\phi^2 - \frac{g}{2\cdot 4!M^2}\phi_0^4\delta\phi^2.
\end{align*}
Then, an equation of motion for $\delta \phi$ is
\begin{align*}
    \Big(\Box - m^2 - \frac{\lambda \phi^2_0}{2} - \frac{g \phi^4_0}{4!M^2}\Big)\delta \phi = 0.
\end{align*}
Introduce the notations:
\begin{align}
\label{D0_D1}
    \mathcal{D}_0 = \Box - m^2 - \frac{\lambda \phi^2_0}{2}, \quad \mathcal{D}_1 = - \frac{g \phi^4_0}{4!M^2}.
\end{align}
The $\hbar$ correction $\Gamma_{\text{eff}}(\phi_0)$ to the classical action
\begin{align*}
    \Gamma_{\text{cl}}(\phi_0) = - i \int d^4x \Big[\frac{1}{2}(\partial_{\mu} \phi)^2+\frac{m^2\phi_0^2}{2} + \frac{\lambda\phi_0^4}{4!} + \frac{g\phi_0^6}{6! M^2}\Big],
\end{align*}
is  \cite{Jackiw:1974cv}:
\begin{align*}
    \Gamma_{\text{eff}}(\phi_0) = \frac{i}{2} \text{Tr} \; \text{ln} (-\mathcal{D}_0 - \mathcal{D}_1),
\end{align*}
(one can also find these textbook calculations, for example, in \cite{Weinberg:1996kr}) or in momentum space
\begin{align*}
    &\Gamma_{\text{eff}}(\phi_0) = \frac{i}{2} \int \frac{d^4k}{(2\pi)^4} \text{ln} (-\mathcal{D}_0 - \mathcal{D}_1), 
\end{align*}
with $\mathcal{D}_0 = -k^2 - m^2 - \frac{\lambda \phi^2_0}{2}$. Next, since we would like to find $K_{\theta,1}$, we need to consider the application of $\theta$-operation on the quantum effective action $\Gamma_{\text{eff}}(\phi_0)$. So, the leading  term in $\Gamma_{\text{eff},\theta}$, which is connected to $\mathcal{D}_0$ operator, is:
\begin{align*}
    \Gamma_{\text{eff},\theta} = -i \frac{d}{dm_0^2}\Big\{ \frac{i}{2} \int \frac{d^4k}{(2\pi)^4}\text{ln}[-\mathcal{D}_0]\Big\},
\end{align*}
and next, we evaluate, using $\mathcal{D}_0 = -k^2 - m^2 - \frac{\lambda \phi^2_0}{2}$ 
\begin{align}
\label{Green}
    &\frac{d}{dm_0^2} \text{ln}\Big[k^2 + m^2 + \frac{\lambda \phi^2_0}{2}\Big] = \frac{1}{k^2 + m^2 + \frac{\lambda \phi^2_0}{2} }\big(1+ \mathcal{O}(\hbar)\big)\equiv G,
\end{align}
where we introduce the factor $1 + \mathcal{O}(\hbar)$ to show that we work in the leading by $\hbar$ order; in other words, this factor comes from $\partial m^2/\partial m^2_0$ and $\partial \lambda/\partial m^2_0$; next, we introduce Green function $G$ for  $\mathcal{D}_0$ operator. Finally, we arrive
\begin{align*}
    \Gamma_{\text{eff},\theta} = \frac{1}{2}\int \frac{d^4k}{(2\pi)^4} G.
\end{align*}
To evaluate the latter, it is convenient to consider \textit{adiabatic expansion} \cite{Birrell,DeWitt} of effective action, counting the number of derivatives of $\phi_0$-field. Firstly, for the simplicity we introduce $\Phi = \lambda \phi_0^2/2$ and then expand it with respect to small $x_{\mu}$:
\begin{align*}
    \Phi = \Phi(0) + \partial_{\mu}\Phi \cdot x^{\mu} + \frac{1}{2} \partial_{\mu}\partial_{\nu}\Phi \cdot x^{\mu}x^{\nu} + \ldots.
\end{align*}
In the momentum space, we have $x_{\mu} = i \partial/\partial k^{\mu}$, and in this representation, one can write an equation for the Green function \eqref{Green} up to $x^{\mu}x^{\nu}$:
\begin{align}
\label{Green_eq}
    &\Big(k^2 + m^2 + \Phi(0)+ \partial_{\mu}\Phi \cdot i \frac{\partial}{\partial k^{\mu}} -  \partial_{\mu}\partial_{\nu}\Phi \cdot \frac{\partial}{\partial k^{\mu}}\frac{\partial}{\partial k^{\nu}}\Big)G = 1.
\end{align}
Since $x_{\mu} = i \partial/\partial k^{\mu}$ is a small parameter, we can use the perturbation theory and power-counting with respect to $x_{\mu} \sim \epsilon$ and find $G$ in the form
\begin{align*}
    G = G_0 + \epsilon G_1 + \epsilon^2 G_2 + \mathcal{O}(\epsilon^3).
\end{align*}
We also use the fact that in \eqref{Green_eq} the term $\partial_{\mu}\Phi \cdot i \frac{\partial}{\partial k^{\mu}}$ is of $\epsilon$ order and the term $ \partial_{\mu}\partial_{\nu}\Phi \cdot \frac{\partial}{\partial k^{\mu}}\frac{\partial}{\partial k^{\nu}}$ is $\sim \epsilon^2$. So, the standard and straightforward calculations lead to
\begin{subequations}
\begin{align}
    &G_0 = \frac{1}{k^2 + m^2 + \Phi},\\
    &G_1 = - \frac{i}{2} \partial_{\mu}\Phi \frac{\partial}{\partial k^{\mu}} G_0^2,\\
    &G_2 = -\frac{1}{2} G_0 \frac{\partial}{\partial k^{\mu}}\frac{\partial}{\partial k^{\nu}} G_0^2 \cdot   \partial_{\mu}\Phi \partial_{\nu}\Phi \nonumber
    \\
    &-\frac{1}{2} G_0 \frac{\partial}{\partial k^{\mu}}\frac{\partial}{\partial k^{\nu}} G_0 \cdot   \partial_{\mu}\partial_{\nu}\Phi,
    \label{G2}
\end{align}
\end{subequations}
and this procedure can be continued up to arbitrary order in $\epsilon$. However, originally we are after the one-loop correction to $K_{1,\theta}$ (the effective kinetic term with one theta operation), i.e. we need to consider the terms proportional to $(\partial_{\mu}\phi_0)^2$. This is the $G_2$ expression \eqref{G2}, so\footnote{In order to make whole expression \eqref{G2} be proportional to $(\partial_{\mu}\phi_0)^2$, one should proceed the integration by parts for the second term in \eqref{G2}, i.e.:
\begin{align*}
    -\frac{1}{2} G_0 \frac{\partial}{\partial k_{\mu}}\frac{\partial}{\partial k_{\nu}} G_0 \cdot   \partial_{\mu}\partial_{\nu}\Phi \to \frac{1}{2} \partial_{\nu}\Phi\partial_{\mu}\Big[G_0 \frac{\partial}{\partial k_{\mu}}\frac{\partial}{\partial k_{\nu}} G_0 \Big].     
\end{align*}}
\begin{align*}
    &K_{1,\theta} = \frac{1}{2} \int \frac{d^4 k }{(2\pi)^4} G_2=\frac{1}{2} \int \frac{d^4 k }{(2\pi)^4} \Big(-\frac{1}{2} G_0 \frac{\partial}{\partial k^{\mu}}\frac{\partial}{\partial k^{\nu}} G_0^2 \cdot   \partial_{\mu}\Phi \partial_{\nu}\Phi+\frac{1}{2} \partial_{\nu}\Phi\partial_{\mu}\Big[G_0 \frac{\partial}{\partial k_{\mu}}\frac{\partial}{\partial k_{\nu}} G_0 \Big]\Big).
\end{align*}
After some algebra one arrives to
\begin{align}
    &K_{1,\theta} = \frac{ i}{(4\pi)^{2}}\frac{(\partial_{\mu}\phi_0)^2}{2}\Big(\frac{\lambda}{2m^2}\Big)^2   \frac{\phi_0^2}{(1 + \frac{\lambda \phi_0^2}{2m^2})^2}.
    \label{K1t_answ}
\end{align}
We note, that the latter expression is valid up to $\big(1 + \mathcal{O}(\hbar)\big)$ order, see the definition \eqref{Green}.
Thus, what we have found from \eqref{K1t_answ} is that $K_{1,\theta} \propto \phi_0^2$ in the leading order by $\lambda$. It means that the second and the third terms in \eqref{CS_K_h} are both proportional to $\phi_0^2$, and we conclude that
\begin{align*}
    \gamma_1 = 0.
\end{align*}
Inserting the latter back to \eqref{CS_K_h}, one can evaluate the one-loop correction to effective kinetic term. Finally, we note that $K_1$ is determined up to an arbitrary constant, which can be fixed by the appropriate boundary condition.

All contributions to $\Gamma_{\theta\theta}$ are finite and are shown in Fig.~\ref{fig:phi_eff}. We recall that $\theta$-operation can be presented as a cutting of a propagator in two and pasting together by $\theta$-vertex, which also brings $(-1)$ into the analytical expressions \cite{Mooij:2021ojy,Mooij:2021lbc}. The corresponding formula for $\Gamma_{\theta\theta}$ in $\hbar$ order reads
\begin{align*}
    \Gamma_{\theta\theta,1} = -\frac{1}{32\pi^2} \text{ln} \Big[1 + \frac{\lambda \phi^2_0}{2m^2} 
+ \frac{g}{4!M^2}\frac{\phi_0^4}{m^2} \Big],
\end{align*}
which is obtained after the summation of all one-loop contributions. In the latter formula, we suppose that we subtract the bubble contributions, which are connected with the cosmological constant (there is a detailed discussion of this topic in Ref.~\cite{Mooij:2024rys}, see Appendix A there). Begin with the first equation \eqref{CS_V_1}; at $\hbar$ order it is given by
\begin{align}
\label{CS_V_1_h}
    \frac{ \Gamma_{\theta,1} }{m^2} = i\Big\{  \frac{G_1 \phi_0^2}{2m^2}
+ \frac{\phi_0^6 \Gamma'_1}{m^6} - \frac{\beta_1 \phi_0^4}{48m^4} 
- \frac{\phi_0^6 \Omega_{g,1}}{1440m^4M^2}\Big\},
\end{align}
where prime means the derivative with respect $\frac{\phi_0^2}{m^2}$. So, using $x$ variable  \eqref{x},  the eq.~\eqref{CS_V_1_h} transforms to
\begin{align}
\label{Gamma_t_1}
    \frac{ \Gamma_{\theta,1} }{m^2} = i\Big\{  \frac{xG_1}{2}+ x^3 \Gamma'_1(x) -\frac{x^2\beta_1}{48} 
- \frac{x^3 m^2\Omega_{g,1}}{1440M^2}\Big\},
\end{align}
where prime now stands for the derivative with respect to $x$.
Then evaluate the equation \eqref{CS_V_2} in $\hbar$ order with the use of result \eqref{Gamma_t_1}:
\begin{align*}
    &x^4 \Gamma''_1(x) + 2 x^3 \Gamma'_1(x) - \frac{x^2 \beta_1}{48} + \frac{x \gamma_{\theta,1}}{4} 
- \frac{x^3 m^2\Omega_{g,1}}{1440M^2} = \frac{1}{32\pi^2} \text{ln} \Big[1 + \frac{\lambda x}{2} 
+ \frac{g x^2 m^2}{4!M^2} \Big],
\end{align*}
and solving the latter one, we arrive at the following answer for one loop correction to the effective potential in $\hbar$ order
\begin{align*}
&\Gamma_1=\frac{1}{64 \pi^2}\left\{  \frac{\text{ln} [1+\frac{x}{24}(12\lambda  
+ g \cdot  x\cdot m^2/ M^2)]}{x^2}+ \text{ln} [24 +x(12\lambda  
+ g \cdot  x\cdot m^2/ M^2)]\left(\frac{\lambda}{x}+\frac{\lambda^2}{4} 
+ \frac{g m^2}{12M^2}\Big(1 + \frac{\lambda x}{2}\Big)\right)
\right\} \nonumber\\
& +\text{ln} [x] \cdot\left[\frac{1}{4 x}\left(\gamma_{\theta, 1}
-\frac{\lambda}{16 \pi^2}\right)+\frac{1}{48}\left(\beta_1-\frac{3 \lambda^2}{16 \pi^2}\right) 
- \frac{g}{768M^2}\Big(\frac{m^2}{\pi^2} + \frac{\lambda x m^2}{2\pi^2} \Big) 
+  \frac{x m^2 \Omega_{g,1}}{1440 M^2}\right] \nonumber\\
& +\frac{1}{4 x}\left(\gamma_{\theta,1}-\frac{\lambda}{32 \pi^2}
-\frac{g\cdot x \cdot m^2}{384\pi^2M^2}+c_1\right)+c_2,
\end{align*}
where $c_1$ and $c_2$ are the integration constants. To obtain the final answer, we should also define $\gamma_{\theta,1}$,  $\beta_1$ and $\Omega_{g,1}$. To this end, we require that our result satisfies the \textit{analyticity requirement}. In other words, we impose that the solution for $\Gamma_1$ is regular with $\phi_0 = 0$ (what guarantees that Green functions exist perturbatively, see the discussion in Ref.~\cite{Mooij:2024rys}). The latter rules the terms proportional to $\ln[x]$ out. This equips us with 
\begin{subequations}
\begin{align}
    &\gamma_{\theta,1} = \frac{\lambda}{16\pi^2},\\
    &\beta_1 = \frac{3\lambda^2}{16\pi^2} + \frac{g m^2}{16\pi^2M^2},\label{beta_our}\\
    &\Omega_{g,1} = \frac{15g \lambda}{16\pi^2},\label{omega_our}
\end{align}
\end{subequations}
and our regular result then reads
\begin{align*}
&\Gamma_1=\frac{1}{64 \pi^2}\Big\{  \frac{\text{ln} [1+\frac{x}{24}(12\lambda  
+ g \cdot  x\cdot m^2/ M^2)]}{x^2}+ \text{ln} [24 +x(12\lambda  
+ g \cdot  x\cdot m^2/ M^2)]\left(\frac{\lambda}{x}+\frac{\lambda^2}{4} + \frac{g m^2}{12M^2}\Big(1 
+ \frac{\lambda x}{2}\Big)\right)
\Big\} \nonumber\\
& +\frac{1}{4 x}\left(\frac{\lambda}{32 \pi^2}
-\frac{g\cdot x \cdot m^2}{384\pi^2M^2}+c_1\right)+c_2.
\end{align*}
The integration constants $c_1$ and $c_2$ can be found by imposing appropriate boundary conditions at some convenient field value. The choice of $c_1$ and $c_2$, or in other words, the choice of boundary conditions, actually defines the physical parameters $m$, $\lambda$, and $g$.

The expressions for $\beta$ functions can be compared with the more general results derived in \cite{Manohar:2024xbh} for a $n$-component real scalar field in two-loop approximation with the use of MSbar scheme based on dimensional regularisation. While reducing the formulas of \cite{Manohar:2024xbh}  to one-loop order and making the same choice of $\{C_1, C_2, C_3\}$ we found a perfect coincidence, see Appendix \ref{app:monohar_compare}.

\subsection{Calculation of correlation functions}
\label{sec:green}
 
The CS method for $n$-point correlation functions contains the following \textit{finite} ingredients: $i)$ convergent connected diagrams; $ii)$ a set of CS equations between $n$-point functions and their derivatives with respect to the mass parameter and; $iii)$ the boundary conditions to fix integration constants or to define parameters from the Lagrangian. Below, we use all these ingredients to find two-, four- and six-point functions at one loop level. 

We begin with the tree contributions to two-, four-, and six-point correlation functions, which can be found from the Lagrangian \eqref{L}:
\begin{subequations}
\label{tree}
\begin{align}
    &\Bar{\Gamma}^{(2)}
 =  i(m^2 + k^2)\,,\\
 &\Bar{\Gamma}^{(4)}
 =  -i \lambda\,,\\
 &\Bar{\Gamma}^{(6)}
 =   -\frac{ig}{M^2} .
\end{align}
\end{subequations}

\begin{figure}
    \centering
\includegraphics[scale=0.43]{
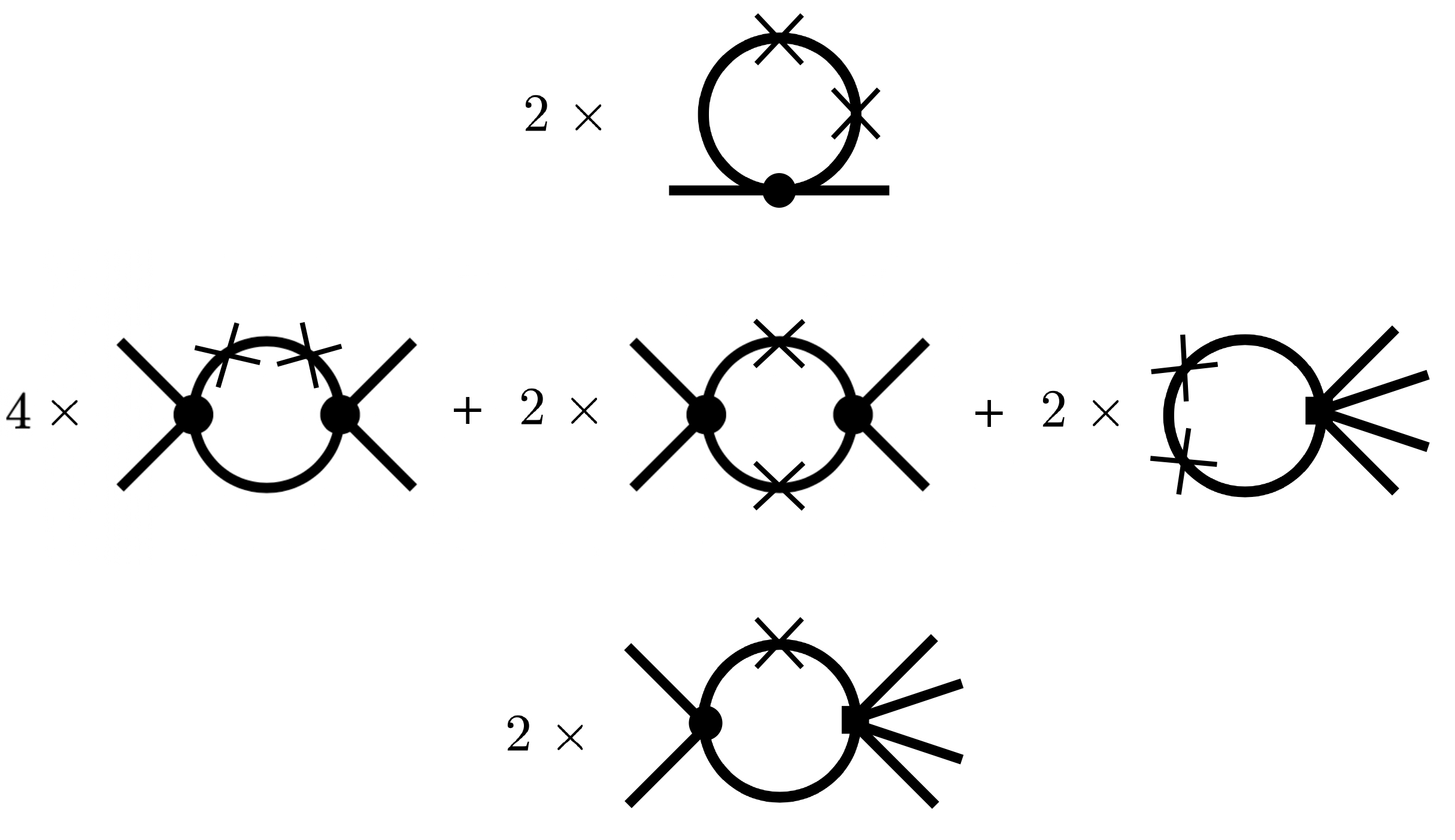}
    \caption{Graphs for 2-, 4-, and 6-point correlation functions with one (for a six-point function) and two (for two- and four-point functions) $\theta$-operations.  The square vertex corresponds to the $\phi^6$ term.}
    \label{fig:diagr}
\end{figure}

The corresponding one-loop Feynman diagram for the 2-point function with all needed theta-operations is shown in Fig. \ref{fig:diagr}, the top one.
The expression for this diagram is
\begin{align}
\label{G2_tt}
    \Bar{\Gamma}^{(2)}_{\theta\theta} = -\frac{i \lambda}{32 \pi^2m^2}.
\end{align}
Next, the one-loop Feynman diagrams for the 4-point function with two theta-operations are shown in Fig. \ref{fig:diagr}, middle line, and the analytical expressions for these graphs are
\begin{align}
\label{G4_tt}
    \Bar{\Gamma}_{\theta\theta}^{(4)}=   
- \frac{i\lambda^2}{2(4\pi)^2}\sum_{3\;\text{opt}}\int^1_0 \frac{dx }{\Delta^2} 
+ \frac{ig}{32 \pi^2 m^2M^2},
\end{align}
where $\Delta \equiv m^2+\kappa_i^2(1-x)x$, with $\kappa_i = \{k_1+k_2,k_1-k_3,k_1-k_4\}$ being the sum of incoming and outgoing momenta in three different $s-,t-,u-$channels, respectively. Here, $x$ is a Feynman parameter as well. Finally, for the six-point correlation function in one loop approximation, we need only one $\ theta$ operation to obtain convergent expression. Indeed, the graph from Fig.~\ref{fig:diagr} (bottom one) corresponds to
\begin{align}
\label{G6_t}
    &\Bar{\Gamma}^{(6)}_{\theta} = 
    -\frac{g\lambda}{2(4\pi)^2M^2} \sum_{n=1}^{15} \int^1_0 \frac{dy   }{ \Delta_6} ,
\end{align}
where $\Delta_6 \equiv m^2 + s_n^2 (1-y)y$, with $s^2_n$ being the combinations of momenta in different 15 channels and $y$ is the Feynman parameter.

Next, these $\Bar{\Gamma}^{(2)}_{\theta\theta}$, $\Bar{\Gamma}^{(4)}_{\theta\theta}$, and $\Bar{\Gamma}^{(6)}_{\theta}$ are used when solving the following CS equations:
\begin{subequations}
\label{CS_corr_func}
\begin{align}
    \label{CS_G2_2}
    &2m^2iG\Bar{\Gamma}^{(2)}_{\theta\theta } = \Big[\left(2m^2\frac{\partial}{\partial m^2} 
+ \beta\frac{\partial}{\partial \lambda} +\Omega_{g}\frac{\partial}{\partial g}\right)+2\gamma+\gamma_{\theta}\Big]\Bar{\Gamma}^{(2)}_{\theta},\\
    \label{CS_G2_1}
    &2m^2iG\Bar{\Gamma}^{(2)}_{\theta} = \Big[\left(2m^2\frac{\partial}{\partial m^2} 
+ \beta\frac{\partial}{\partial \lambda} +\Omega_{g}\frac{\partial}{\partial g}\right)+2\gamma\Big]
\Bar{\Gamma}^{(2)},\\
    \label{CS_G4_2}
    &2m^2iG\Bar{\Gamma}^{(4)}_{\theta\theta } = \Big[\left(2m^2\frac{\partial}{\partial m^2} 
+ \beta\frac{\partial}{\partial \lambda} +\Omega_{g}\frac{\partial}{\partial g}\right)+4\gamma+\gamma_{\theta}\Big]\Bar{\Gamma}^{(4)}_{\theta},\\
    \label{CS_G4_1}
    &2m^2iG\Bar{\Gamma}^{(4)}_{\theta} = \Big[\left(2m^2\frac{\partial}{\partial m^2} 
+ \beta\frac{\partial}{\partial \lambda} +\Omega_{g}\frac{\partial}{\partial g}\right)
+4\gamma\Big]\Bar{\Gamma}^{(4)},\\
    \label{CS_G6_1}
    &2m^2iG\Bar{\Gamma}^{(6)}_{\theta} = \Big[\left(2m^2\frac{\partial}{\partial m^2} 
+ \beta\frac{\partial}{\partial \lambda} +\Omega_{g}\frac{\partial}{\partial g}\right)
+6\gamma\Big]\Bar{\Gamma}^{(6)},
\end{align}
\end{subequations}
which are nothing but the equations on $\Bar{\Gamma}^{(2)}$, $\Bar{\Gamma}^{(4)}$, and $\Bar{\Gamma}^{(6)}$.  As we have commented earlier, these equations directly follow from the general CS equation \eqref{general}. The parametrisation we use is as follows:
\begin{subequations}
\label{param_Gammas}
\begin{align}
    &\Bar{\Gamma}_{\theta\theta}^{(n)}  = \Bar{\Gamma}^{(n)}_{\theta\theta,0} 
+\hbar \cdot \Bar{\Gamma}_{\theta\theta,1}^{(n)} + \mathcal{O}(\hbar^2) ,\\
     &\Bar{\Gamma}_{\theta}^{(n)}  = \Bar{\Gamma}^{(n)}_{\theta,0} 
+\hbar \cdot\Bar{\Gamma}_{\theta\theta,1}^{(n)} + \mathcal{O}(\hbar^2),\\
     &\Bar{\Gamma}^{(n)}  = \Bar{\Gamma}^{(n)}_{0} +\hbar\cdot \Bar{\Gamma}_{1}^{(n)} 
+ \mathcal{O}(\hbar^2),
\end{align}
\end{subequations}
and 
\begin{subequations}
\label{param_betas}
\begin{align}
    &G = G_{0}+\hbar\cdot G_{1}+\mathcal{O}(\hbar^2),\\
    &\gamma = \gamma_{0}+\hbar\cdot\gamma_{1}+\mathcal{O}(\hbar^2),\\
    &\gamma_{\theta} = \gamma_{\theta,0}+\hbar\cdot\gamma_{\theta,1}
+\mathcal{O}(\hbar^2),\\ 
    &\beta = \beta_0 + \hbar\cdot \beta_1 +\mathcal{O}(\hbar^2),\\
    &\Omega_g = \Omega_{g,0} + \hbar\cdot \Omega_{g,1} +\mathcal{O}(\hbar^2)\,,
\end{align}
\end{subequations}
So everything is written in the same manner as in the effective potential consideration, see Sec. \ref{sec:eff_poten}.  At $\hbar^0$ order we have \eqref{tree}, so we can legitimately write 
\begin{subequations}
\label{tree_theta}
\begin{align}
    &\Bar{\Gamma}^{(2)}_{\theta,0} = 1, \quad \Bar{\Gamma}^{(2)}_{\theta\theta,0} = 0,\\
    &\Bar{\Gamma}^{(4)}_{\theta,0} = 0, \quad 
    \Bar{\Gamma}^{(4)}_{\theta\theta,0} = 0,\\
    &\Bar{\Gamma}^{(6)}_{\theta,0} = 0.
\end{align}
\end{subequations}
Inserting the latter together with \eqref{param_Gammas} and \eqref{param_betas} into all CS equations \eqref{CS_corr_func} and keeping only $\hbar^0$ terms, one arrives to:
\begin{align*}
    (G_0 -1 - \gamma_0)m^2 - \gamma_0 k^2 = 0, \quad &\text{from} \quad \text{Eq.~\eqref{CS_G2_1}},\\
    \beta_0 + 4 \gamma_0 \lambda = 0, \quad &\text{from}  \quad \text{Eq.~\eqref{CS_G4_1}},\\
    6 g \gamma_0 + \Omega_{g,0}  = 0, \quad &\text{from} \quad \text{Eq.~\eqref{CS_G6_1}},\\
    2 \gamma_0 + \gamma_{\theta,0}  = 0, \quad &\text{from} \quad \text{Eq.~\eqref{CS_G2_2}},
\end{align*}
what defines all zeroth order parameters as 
\begin{align}
\label{san_1}
    G_0 = 1, \;\;\; \gamma_0 = 0, \;\;\;  \beta_0 = 0,   \;\;\; \Omega_{g,0}  =  0,
 \;\;\;  \gamma_{\theta,0} = 0.
\end{align}
Note, that \eqref{CS_G4_2} is satisfied automatically at \eqref{san_1} set in $\hbar^0$ order.

Moving forward, we turn to the $\hbar$ order. Begin with the Eq.~\eqref{CS_G4_2} where we also substitute \eqref{G4_tt}, so
\begin{align*}
   2 i m^2  \Big(- \frac{i\lambda^2}{2(4\pi)^2}\sum_{3\;\text{opt}}\int^1_0 \frac{ dx }{ \Delta^2} 
+ \frac{ig}{32 \pi^2 m^2 M^2}\Big)= 2 m^2 \frac{\partial}{\partial m^2}\Bar{\Gamma}^{(4)}_{\theta,1},
\end{align*}
and the solution is
\begin{align*}
    &\Bar{\Gamma}^{(4)}_{\theta,1} = b_1 
+ \frac{g}{32 \pi^2 M^2} \frac{1}{3}\Big(\text{ln}\Big[\frac{s}{m^2} \Big]
+\text{ln}\Big[\frac{t}{m^2}  \Big]+\text{ln}\Big[\frac{u}{m^2}  \Big] \Big)-  \frac{\lambda^2}{32 \pi^2 }\sum_{3\;\text{opt}}\int^1_0 \frac{ dx }{\Delta},
\end{align*}
where $b_1$ is some dimensional constant of integration. We substitute this answer into \eqref{CS_G4_1}, solve it and arrive to
\begin{align*}
    \Bar{\Gamma}^{(4)}_1 &= b_2 + \frac{i}{96 \pi^2} \Big\{  \frac{3g m^2}{M^2} 
+ 96 b_1 m^2 \pi^2 +\Big(\text{ln}\Big[\frac{s}{m^2} \Big]
+\text{ln}\Big[\frac{t}{m^2}  \Big]+\text{ln}\Big[\frac{u}{m^2}  \Big] \Big)\Big(-16 \pi^2 (\beta_1 + 4\lambda \gamma
    _1) + \frac{g m^2}{M^2} + 3 \lambda^2\Big) \nonumber\\ 
&-  3\lambda^2\sum_{3\;\text{opt}}\int^1_0 dx 
\cdot \text{ln}\frac{\Delta}{m^2}\Big\}.
\end{align*}
Imposing that the solution is regular at $\kappa_i^2\to 0$ (i.e. terms with log are forbidden), we find 
\begin{align*}
   -16 \pi^2 (\beta_1 + 4\lambda \gamma
    _1) + \frac{g m^2}{M^2} + 3 \lambda^2 = 0, 
\end{align*}
so
\begin{align*}
    \beta_1 + 4\lambda \gamma
    _1 = \frac{3 \lambda^2 + \frac{g m^2}{M^2}}{16 \pi^2}.
\end{align*}
Then, the regular answer for the 4-point function at $\hbar$ order is
\begin{align*}
    &\Bar{\Gamma}^{(4)}_1 = b_2 + i b_1 m^2 +   \frac{i g m^2}{32\pi^2 M^2}   
-  \frac{i\lambda^2}{32\pi^2}\sum_{3\;\text{opt}}\int^1_0 dx \cdot \text{ln}\frac{\Delta}{m^2}.
\end{align*}
Two integration constants $b_1$ and $b_2$ can be defined from 
boundary conditions at a chosen value of momenta.

Evaluating equation \eqref{CS_G2_2} with \eqref{G2_tt} we arrive to
\begin{align*}
    \Bar{\Gamma}^{(2)}_{\theta,1} = b_3 
+ \frac{\text{ln}\big(\frac{k^2}{m^2}\big)}{32\pi^2}\big(16\pi^2(2\gamma_1 
+ \gamma_{\theta,1})-\lambda\big),
\end{align*}
with dimensionless $b_3$. Again, we require the regular behaviour, so
\begin{align*}
    2\gamma_1 + \gamma_{\theta,1} = \frac{\lambda}{16\pi^2}.
\end{align*}
The equation \eqref{CS_G2_1} gives
\begin{align*}
    \Bar{\Gamma}^{(2)}_{1} = b_4 + i\Big(m^2 (b_3 + G_1 - \gamma_1) 
+ k^2 \gamma_1 \text{ln}\frac{k^2}{m^2}\Big),
\end{align*}
so
\begin{align*}
    \gamma_1 = 0.
\end{align*}
The regular answer for the 2-point correlation function is
\begin{align*}
    \Bar{\Gamma}^{(2)}_{1} = b_4 + im^2 (b_3 + G_1 ),
\end{align*}
with two integration constants $b_3$ and $b_4$. The parameter $G_1$ comes together with $b_3$ everywhere, so this $G_1$ can be just absorbed into $b_3$. Finally, find six-point correlation function from CS equation \eqref{CS_G6_1}, using \eqref{G6_t}
\begin{align*}
    &\Bar{\Gamma}^{(6)}_{1} = b_5 - \frac{i}{32\pi^2}
\Big( \frac{16 \pi^2 \Omega_{g,1} }{M^2}\sum\text{ln}\Big[\frac{s_n^{ 2}}{m^2}\Big] - \frac{15g\lambda}{M^2}\sum \text{ln}\Big[\frac{s_n^{2}}{m^2}\Big] + \frac{g \lambda}{M^2}\sum_{15\;\text{opt}}\int_0^1 dy\text{ln}\frac{\Delta_6}{m^2}\Big),
\end{align*}
where $\sum\text{ln}\Big[\frac{s_n^{ 2}}{m^2}\Big]$ is the full sum of all momenta combinations for 15 channels. The analyticity provides
\begin{align*}
    \frac{16 \pi^2 \Omega_{g,1} }{M^2} - \frac{15g\lambda}{M^2} = 0,
\end{align*}
thus
\begin{align*}
    \Omega_{g,1} =\frac{15g\lambda}{16\pi^2},
\end{align*}
and six-point function reads
\begin{align*}
    \Bar{\Gamma}^{(6)}_{1} = b_5 - \frac{i}{32\pi^2}  
\frac{g \lambda}{M^2}\sum_{15\;\text{opt}}\int_0^1 dy\cdot 
\text{ln}\frac{\Delta_6}{m^2},
\end{align*}
with $b_5$ being an integration constant.

Let us briefly comment on the results from this section. The answers for $G$, $\gamma$, $\gamma_{\theta}$, $\beta$, and $\Omega_g$ coincide with the results from Sec.~\eqref{sec:eff_poten}, i.e. with the effective potential consideration. We have defined these parameters using the property of analyticity, which is more general than the use of boundary conditions. Nevertheless, the boundary conditions can be used at the final stage of all evaluations to define the integration constants. For example, one can pick
\begin{align*}
    &\Bar{\Gamma}^{(2)}|_{k^2=0}
 =  im^2   ,
\end{align*}
\begin{align*}
    &\Big[\frac{d}{dk^2}\Bar{\Gamma}^{(2)}\Big]_{k^2=0}
 =  i   \,,
\end{align*}
\begin{align*}
    &\Bar{\Gamma}^{(4)}|_{\kappa_i^2=0}
 =  -i \lambda  ,
\end{align*}
which were used in Ref.~\cite{Mooij:2021ojy,Mooij:2021lbc,Callan:1975} and are the definitions of physical mass and coupling constant $\lambda$. Another one can be written as
\begin{align*}
    &\Bar{\Gamma}^{(6)}|_{s_n^2=0}
 =   -\frac{ig}{M^2}  ,
\end{align*}
which is necessary in the case of non-renormalizable theory and also defines the constant $g$.

\section{Conclusion}
\label{sec:summary}
The non-renormalizable theories are of the greatest interest to study. For example, one of the most important theories -- gravity -- manifests itself as a non-renormalizable theory. This motivated us to find the generalisation of the CS method to the case of non-renormalizable theories. In this paper, we have found out that it is possible to write down a unique \textit{generalised CS equation} \eqref{general}, which is formulated in terms of specific functional \eqref{functional}. This equation \eqref{general} unifies all CS equations, i.e. generates them for effective action, for effective potential and any correlation functions in any order by $\hbar$. 

To illustrate how it all works, we choose the specific non-renormalizable model \eqref{L} containing $\phi^6/M^2$  interaction. Using the CS equations, we evaluated the one-loop correction to the effective potential $\Gamma_1$ and  $2$-, $4$-, and $6$-point correlation functions,  keeping the leading terms in $1/M^2$ expansion, and determined the anomalous dimensions and beta-functions. No divergences have been met at any stage of the computations. No fine-tuning of the small mass parameter $m^2$ is needed as well, and there is no impact of the high energy $M^2$-scale physics on the low energy $m^2$-scale one.

So, to summarize, we have suggested a new method for the computation of the physical Green's functions and anomalous dimensions in effective field theories without any regularisation or intermediate infinities and demonstrated how it works in a simple scalar field theory. To the best of our knowledge, the equations \eqref{k_theta}, \eqref{functional}, and \eqref{general}   have never appeared in the literature before. To show how the method works, we provided a detailed \textit{step-by-step} calculation and final answers for physical $2$-, $4$-, and $6$- correlation functions in the non-renormalisable (or effective in another language)  $\phi^4+\phi^6/M^2$ theory in one loop. Thus, the considered method applies to the study of different features in EFT, and the latter is under constant development nowadays, with a lot of research done in the last few years, see, for instance, Refs.~\cite{Manohar:2024xbh,Manohar:2018aog,Cohen:2019wxr,Neubert:2019mrz,Baratella:2020lzz,Cohen:2020xca,Assi:2023zid}, references therein and many others.

The CS method explored in this paper and in  \cite{Mooij:2021ojy,Mooij:2021lbc,Mooij:2024rys}  is only applicable to the {\em massive} fields, leaving aside gauge theories and gravity. It would be interesting to search for the generalisation of the method to a more general class of (non-renormalisable) theories involving the massless fields. Perhaps the ideas expressed in  \cite{Lehmann:1954rq,Nishijima:1960zz,Tkachov:1994gdc,tHooft:2004bkn,Moffat:2010bh,Moffat:2014zwa,Rejzner:2016hdj,Lenshina:2019pyf,Lenshina:2020qdv,Lenshina:2020edt,Kataev:2013eta,Kataev:2017qvk,Kazakov:2020mfp,Moffat:2020ecn,Green:2020rsb,Morgan:2021jpa,Kazakov:2023ugc,Kazakov:2023raj,Kazakov:2023tii,DM:2024uty,Filippov:2024nza,Arefeva:1977bt,Arefeva:1979bd,Arefeva:1978fj} may appear to be helpful for this aim.

\section*{Acknowledgments} 

The authors are grateful to Andrei Kataev, Maxim Libanov, Sergei Demidov, Bulat Farkhtdinov, and Dmitry Ageev for useful comments and fruitful discussions. MS is grateful to Sander Mooij for many invaluable discussions and comments. The work of YA on Sec. \ref{sec:nonrenorm} has been supported by Russian Science Foundation Grant No. 24-72-10110. PP was supported by IBS under the project code IBS-R018-D3. The work of MS was supported by the Generalitat Valenciana grant PROMETEO/2021/083.

\appendix

\section{General expressions for beta functions in one loop}
\label{app:monohar_compare}
\numberwithin{equation}{section}

\setcounter{equation}{0}
In Ref.~\cite{Manohar:2024xbh}, the $\beta$-functions for the $O(n)$ symmetric theory of n-component scalar field $\phi$ with dimension six operators were computed in two-loop approximation with the use of MSbar scheme based on dimensional regularization. In this Appendix, we show that our results agree with those of \cite{Manohar:2024xbh} when they are reduced to one loop and $n=1$.

To this end, let us write down an initial bare (below the subscripts $b$ refer to bare quantities) Lagrangian of Ref.~\cite{Manohar:2024xbh}:  
\begin{align}
\label{Lagr_Manohar}
    &\mathcal{L} =\frac{1}{2}\partial_{\mu}\phi_b\partial^{\mu}\phi_b 
- \frac{1}{2}m^2_b\phi_b \phi_b  - \frac{\lambda_b}{4}(\phi_b \phi_b)^2  + C_{4,b}\mathcal{O}_{4,b}  + D_{4,b}\mathcal{R}_{4,b} + C_{6,b}\mathcal{O}_{6,b} + D_{2,b}\mathcal{R}_{2,b}.
\end{align}
The signature of the metric in \cite{Manohar:2024xbh} is $(+,-,-,-)$ and differs from ours. All operators are defined as
\begin{align}
    &\mathcal{O}_4=\left(\partial_\mu \phi \cdot \partial^\mu \phi\right)(\phi \cdot \phi), \quad \mathcal{R}_4=\left(\phi \cdot \partial_\mu \phi\right)^2 ,\nonumber\\
&\mathcal{O}_6=(\phi \cdot \phi)^3,  \quad\mathcal{R}_2=\left(\partial_\mu \partial^\mu \phi \cdot \partial_\nu \partial^\nu \phi\right).\nonumber
\end{align}

In Ref.~\cite{Manohar:2024xbh}, it was shown that after field redefinitions, the Lagrangian \eqref{Lagr_Manohar} can be written as \cite{Manohar:2024xbh}:
\begin{align}
\label{redef_L}
    &\mathcal{L} =\frac{1}{2}\partial_{\mu}\tilde{\phi}_b\partial^{\mu}\tilde{\phi}_b 
- \frac{1}{2}\bar{m}^2_b\tilde{\phi}_b\tilde{\phi}_b  - \frac{\bar{\lambda}_b}{4}(\tilde{\phi}_b \tilde{\phi}_b)^2  + \bar{C}_{4,b}\tilde{\mathcal{O}}_{4,b}  + \bar{C}_{6,b}\tilde{\mathcal{O}}_{6,b},
\end{align}
where tilde refers to the canonically normalised bare scalar field. The coefficients $D_{2,4}$ of ``redundant'' operators $\mathcal{R}_{2,4}$ can be eliminated with the specific choice of the constants\footnote{The bar over mass, $\lambda$ as well as over $\bar{C}_{4,b}$ and $\bar{C}_{6,b}$ in \eqref{redef_L}  means that one sticks to some concrete choice of the constants characterising field redefinition.} entering the field redefinition. 

The expressions for beta functions in one-loop approximation read \cite{Manohar:2024xbh}
\begin{align}
&16\pi^2\beta_{\bar{\lambda}}=2(n+8) \bar{\lambda}^2-16(n+3) \bar{\lambda} \bar{m}^2 \bar{C}_4-24(n+4) \bar{m}^2 \bar{C}_6,\nonumber\\
&16\pi^2\beta_{\bar{C}_4}=4(n+2)\bar{\lambda} \bar{C}_4,\nonumber\\
&16\pi^2\beta_{\bar{C}_6}=20 \bar{\lambda}^2 \bar{C}_4+6(n+14) \bar{\lambda} \bar{C}_6,\nonumber
\end{align}
where
\begin{align}
\label{beta_manohar}
    \beta_{\bar{C}_i}(\{\bar{C}\}) = \mu \frac{\mathrm{d} \bar{C}_i}{\mathrm{~d} \mu}, \quad \beta_{\bar{\lambda}}(\bar{\lambda}) = \mu \frac{\mathrm{d} \bar{\lambda}}{\mathrm{~d} \mu},
\end{align}
what coincides with our definition \eqref{our_beta_def}.

Now, we can turn to the comparison with our results. For $n = 1$ the operators $\mathcal{O}_4 = \left(\partial_\mu \phi \cdot \partial^\mu \phi\right)(\phi \cdot \phi)$ and $\mathcal{R}_4 =\left(\phi \cdot \partial_\mu \phi\right)^2$ are exactly the same. The latter means that term $\bar{C}_{4,b}\tilde{\mathcal{O}}_{4,b}$ may be excluded (together with $\tilde{\mathcal{R}}_{4,b}$-term), i.e. one can make $\bar{C}_{4,b} = 0$. Bearing in mind all these points, we arrive to
\begin{subequations}
\label{beta_c4_zero}
\begin{align}
\label{a4.a}
&\beta_{\bar{\lambda}}=\left\{18 \bar{\lambda}^2-24\cdot 5 \bar{m}^2 \bar{C}_6\right\}_1,\\
&\beta_{\bar{C}_4}=0,\\
&\beta_{\bar{C}_6}=\left\{6\cdot 15 \bar{\lambda} \bar{C}_6\right\}_1.\label{a4.c}
\end{align}
\end{subequations}
After several straightforward substitutions of $\bar{\lambda}_{\text{Manohar}} \to \lambda_{\text{our}}/6$ as well as $\bar{C}_{6,b} \to -\frac{g}{6! M^2}$ in \eqref{beta_manohar} and \eqref{beta_c4_zero} one can easily find out, that \eqref{a4.a} is in a perfect agreement with \eqref{beta_our} and \eqref{a4.c} coincides with \eqref{omega_our}.

\end{document}